# Perseus Technology: New Trends in Information and Communication Security[*]

Eric Filiol

Laboratoire de virologie et de cryptologie opérationnelles

France

http://sites.google.com/site/ericfiliol

ffiliol@gmail.com

November 11, 2018

**Abstract**

Using cryptography to protect information and communication has bacically two major drawbacks. First, the specific entropy profile of encrypted data makes their detection very easy. Second, the use of cryptography can be more or less regulated, not to say forbidden, according to the countries. If the right to freely protect our personal and private data is a fundamental right, it must not hinder the action of Nation States with respect to National security. Allowing encryption to citizens holds for bad guys as well.

In this paper we propose a new approach in information and communication security that may solve all these issues, thus representing a rather interesting trade-off between apparently opposite security needs. We introduce the concept of scalable security based on computationnally hard problem of coding theory with the Perseus technology.

The core idea is to encode date with variable punctured convolutional codes in such a way that any cryptanalytic attempt will require a time-consuming encoder reconstruction in order to decode. By adding noise in a suitable way, that reconstruction becomes untractable in practice except for Intelligence services that however must use supercomputers during a significant, scalable amount of time. Hence it limits naturally any will to unduly performs such attacks (eg. against citizens' privacy).

On the users' side, encoder and noise parameters are first exchanged through an initial, short HTTPS session. The principles behind that approach have been mathematically validated in 1997 and 2007. We present the Perseus library we have developed under the triple GPL/LGPL/MPL licences. This library can be used to protect any kind of data.

---

[*]This work has been presented at *i*AWACS 2010.

# 1  Introduction

A necessary – but not sufficient – condition for cryptographic security lies in the secret key size. Cryptography is itself defined as the use of a secret quantity – the key – while coding uses open, widely known mathematical objects without any secret quantity.

The main issue is then: can cryptography be characterized by the presence of a secret quantity only? While it is a necessary condition, it is not a sufficient one. The deep and careful analysis of cryptographic laws of most countries (and international organizations) shows that the "legal" definition of what crypto really is and what is not, relates directly to following (noise) probability

$$P[c_t = m_t \oplus e_t] = P[e_t = 1]$$

where $c_t$ and $m_t$ are the ciphertext and plaintext bits respectively and where $e_t$ can be defined as the noise bit produced by the key the cryptosystem[1] (at time instant $t$). Then, if $P[e_t = 1] = \frac{1}{2} \pm \epsilon$ with $\epsilon$ very close to zero, then it is cryptography, otherwise ($\epsilon$ significantly different from 0) it is coding theory. But are differences between cryptography and coding theory so easy to define? Known cryptanalysis techniques intend to deal with the first case more or less efficiently. On the other side, there are a lot of decoding problems that are computationally hard.

In this paper we are going to consider such a computationally hard problem in order to provide a new information and communication protection scheme whose security level is scalable. We have called it PERSEUS[2] technology and we present here the open source library we have developed to protect any kind of data and protocols.

PERSEUS technology's core idea is to encode data with punctured convolutional codes. Those codes are commonly used in telecommunications (GSM, satellite...) due to their very high encoding speed and their high correcting power. After this encoding layer and right before transmission, an artificial noise is applied to the data flow (as would any channel do). The noise is generated according to noise parameter $p = P[e_t = 1]$ where $e_t$ is the noise bit at time instant $t$. The value of $p$ is around 0.3. Since the convolutational encoder is changing very frequently the attacker always has first to reconstruct the encoder in order to be able to decode. This reconstruction has been proven to be a computationally hard problem [3, 6, 7, 8]. By scalable we mean that if it is always possible to break PERSEUS-protected data, the difficulty can be tuned up in order to require more or less computational efforts: from a few days to a few months on a supercomputer. In addition, only an equivalent, non-punctured encoder can be recovered [7]. However this problem may still remain tractable to solve for any intelligence agency with a suitable computing power.

---

[1] This holds also for block ciphers where the "effect" of the key on the plaintext block can formalized in this way.

[2] Perseus is the mythic hero of Greek mythology who killed the Gorgon Medusa [20]. The botnets – against which PERSEUS technology has been designed initially [5] – are themselves often compared to Medusa and its long tentacles.

The different parameters of the variable encoders are randomly generated: polynomial size constraint, encoding rate, matrix puncturing, noise parameter $p$, encoder polynomials... Then a short HTTPS initial session allows to communicate those parameters to the recipient (about 256 bytes). The recipient and only him is able first to get rid of the artificial deterministic noise and then to set up the suitable Virberi algorithm for data decoding.

What the interest of using scalable security while generally only strong, unbreakable cryptography offers real security? Why would users prefer PERSEUS technology instead of strong cryptography? On the other side, why existing national or international regulations would tolerate the use of this technology? There are on the contrary many reasons to favour the PERSEUS approach over strong encryption.

- The use of encryption, besides the fact that it would lead to severe constraints (encryption overhead, key management...) poses problems in terms of legal regulations, especially in the context of transnational streams with respect to the different national regulations. Then a critical issue arises: how can we protect our personal and private data while still allowing the necessary action of States (for national security for instance) in the field of communication surveillance and whithout lessening the transmission rate significantly? Scalable security offered by PERSEUS provides such a trade-off very efficiently. Any PERSEUS-protected data can be broken provided that a significant amount of time of supercomputer is spent. This limits any States' intents to spy innocent people not involved in terrorism, mafia activities, child pornography... and making them focusing on really bad guys. Moreover the generalization of encryption is not a good thing as pointed out by the US *National Security Agency* [16, 14] and British MI-5 [11] about HADOPI's French questionable initiative. Favouring the use of encryption to protect illegal downloading can severely hinder the cryptanalysis activities of States for national security purposes.

- Why use noisy encoded data instead of encrypted data? Encrypted data by nature exhibit a maximal entropy profile. It is then easy to detect encrypted data. On the contrary, noisy encoded data can exhibit a lower entropy profile which remains closer to that of plain, unencoded data. This lower statistical profile enables to bypass any detection by entropy test or any other statistical detection while encrypted data do not.

To summarize these two strong points of PERSEUS technology, let us consider an illustrative example. John Doe is a US journalist in China. He wants to send a serie of papers about China's Human Rights infringements (and about the 2010 Chinese Peace Nobel Prize). Sending his papers to his agency in USA would be blocked by Chinese authorithies whenever encrypted. On the contrary, using PERSEUS will require a significant time to detect (due to the low entropy profile) and to break. The journalist will have time to go back and safe to USA.

This paper is organized as follows. Section 2 recalls basic facts about convolutional codes and their reconstruction. Section 3 presents the PERSEUS library structure while Section 4 deals with its detailed implementation. Section 5 presents the different experimental results we have obtained with respect to final data entropy and performance while Section 7 concludes by considering future evolution of this library.

## 2 Theoretical Background

In this section, we are going to recall what a (punctured or not) convolutional code is as well as the main results with respect to their reconstruction. The aim is just to provide the reader with the required background to understand the interest of those codes and why they are particularly suitable for our approach. The interested reader will refer to [13] for a more detailed presentation on convolutional codes.

### 2.1 Convolutional Codes

A convolutionnal encoder can be seen as an encoding system (based on a set of $k$ shift-registers without feedback) such that, at each time instant, $k$ information digits (typically the bits of data) enter the encoder (one per register). Each information digit remains in the encoder for $K$ time units and may affect each output during that time. The constant $K$ is the constraint length or the *memory* of the encoder.

At each time instant, $n$ information digits are output, each of them resulting from the XOR of $k$ digits produced by the action of $n$ polynomials on each register. The encoder is thus said to be of rate $\frac{k}{n}$. The action of the $kn$ polynomials and the shift are easily described by polynomial multiplications [8]. So the polynomial representation will be used to represent the different streams.

A message will be composed of $k$ interlaced input streams, each of them represented as a polynomial of degree $N+t$ denoted $a_i(x)$, $i = 1, \ldots, k$. The $kn$ polynomials are of degree $N$ (hence $N = K - 1$) and will be noted $f_{i,j}(x)$. Then the encoder produces $n$ output streams (of length $t$) represented as polynomials of degree $t$, $c_j(x)$, $j = 1, \ldots, n$ and we then have:

$$\sum_{i=1}^{k} a_i(x) f_{i,j}(x) = u_{j,1}(x) + x^N c_j(x) + x^{N+t} u_{j,2}(x) \tag{1}$$

The polynomials $u_{j,1}(x)$ (resp. $u_{j,2}$) (the filling (resp. the emptying) of the registers) are of degree at most $N - 1$. Then the coded sequence is composed of the $n$ interlaced output streams.

Thus the parameters of a convolutionnal encoder are:

- $k$ and $n$ defining the rate and the number of polynomials,

- $K$ the constraint length (in fact it is related to internal memory of the encoder),

- the $kn$ polynomials $f_{i,j}(x)$ of degree $N = K - 1$.

The convolutionnal encoder then describes a $(n, k, N)$-code. Generally, $n$ and $k$ are small integers with $k < n$. The most frequent case is $k = n - 1$. On the contrary, $N$ must be made large enough to achieve low residual decoding error probabilities. The symbols are usually elements of $GF(2)$ but generalization to $GF(q)$ where $q$ is some prime power ($q = p^m$ for some positive integer $m$) can be easily done. We will only consider the case $q = 2$ but all the implementation and results can be generalized to any other prime $q$. This could be interesting in increasing the encoding speed.

Figure 1 describes a convolutional encoder of rate $\frac{1}{2}$.

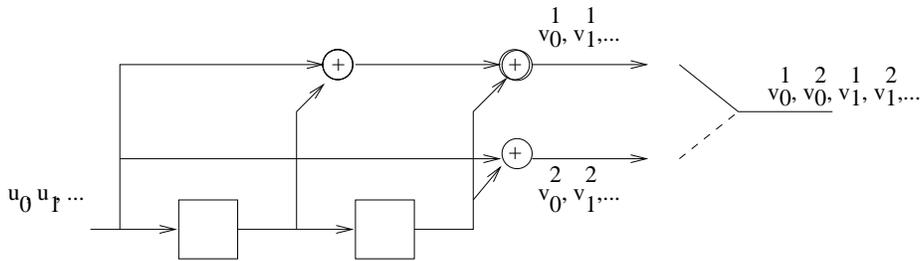

Figure 1: Convolutional encoder of rate $\frac{1}{2}$

In the context of PERSEUS, we will add an artificial noise of parameter $p$ to the (encoded) output sequence $\mathbf{v} = v_0^{(1)}, v_0^{(2)}, v_1^{(1)}, v_1^{(2)}, \ldots$

The decoding step is performed through the classical Viterbi algorithm whose complexity is exponential in $k.N$. Hence, generally their use is limited to codes of short lengths and to reduced encoding rate $\frac{k}{n}$. However in our case since we completely master the noise (we exactly know where the noise bits are applied while any botnet agent does not), we can work with far higher values.

## 2.2 Punctured Convolutional Codes

Punctured convolutional codes were introduced by Cain *et al.* [4] as means of greatly simplifying both Viterbi and sequential decoding of high rate convolutional codes at the expanse of a relatively small performance penalty.

A punctured convolutional code $\mathcal{C}$ is obtained by periodically deleting output symbols from a (base) $(n, k, N)$-convolutional code $\mathcal{C}_b$. Output symbols from $\mathcal{C}_b$ are deleted according to a periodic puncturing pattern (or perforation pattern) which can be described by its puncturing matrix:

$$P = \begin{bmatrix} p_{1,1} & \cdots & p_{1,M} \\ \vdots & & \vdots \\ p_{n,1} & \cdots & p_{n,M} \end{bmatrix}$$

A very important problem is that of the reconstruction of such codes (punctured or not). In an attack context, a monitor wants to have access to the transmitted information (*the message*) without any knowledge on the encoder which produces the intercepted stream (*the coded sequence*). The only way is to reconstruct the encoder, that is to say to recover all its parameters. A simple decoding then gives access to the message provided that the channel noise is not too high (less than a very few percents).

Let us consider a $(n, k, N)$-(base) convolutional code $\mathcal{C}_b$. A given puncturing pattern $P$ is a $n \times M$ $0-1$ matrix with a total of $I$ 1's and $nM - I$ 0's where $p_{i,j} = 0$ indicates that the i-th symbol of every branch in the j-th treillis section (of the treillis diagram of $\mathcal{C}_b$) is to be deleted.

Then the original code $\mathcal{C}_b$, after being punctured with pattern $P$, has become a $(I, kM, m)$-(punctured) code [3] $\mathcal{C}$ [15].

Let us consider an illustrative, simple example.

**Example 1** *Let us take the $(2, 1, 3)$ code with polynomials*

$$(1 + x^2, 1 + x + x^2)$$

*The two output streams can be denoted as follows:*

$$\begin{pmatrix} x_0 & x_1 & x_2 & x_3 & x_4 & x_5 & \ldots \\ y_0 & y_1 & y_2 & y_3 & y_4 & y_5 & \ldots \end{pmatrix}$$

*By using the following puncturing pattern:*

$$P = \begin{pmatrix} 1 & 0 \\ 1 & 1 \end{pmatrix}$$

*we then obtain the two following output streams:*

$$\begin{pmatrix} x_0 & & x_2 & & x_4 & & \ldots \\ y_0 & y_1 & y_2 & y_3 & y_4 & y_5 & \ldots \end{pmatrix}$$

*that we can rearrange as follows:*

$$\begin{pmatrix} x_0 & x_2 & x_4 & \ldots \\ y_0 & y_2 & y_4 & \ldots \\ y_1 & y_3 & y_5 & \ldots \end{pmatrix}$$

*It becomes then obvious that this puncturing produces a new encoder producing three output streams.*

*By use of polycyclic pseudo-circulant matrices [7], the new parameters are easily defined and we have the 6 following polynomials*

$$f_{1,1}(x) = 1 + x \quad f_{1,2}(x) = 1 + x \quad f_{1,3}(x) = 1$$

$$f_{2,1}(x) = 0 \quad f_{2,2}(x) = x \quad f_{2,3}(x) = 1 + x$$

*where $f_{i,j}$ denotes the j-th parity-check polynomial applied on input message stream i.*

As for PERSEUS is concerned, the puncturing pattern $P$ is the last parameter to exchange during the initial HTTPS session.

---

[3]In fact, the degree of the punctured code may be less than $N$, but for most interesting punctured codes no degree reduction will take place

## 2.3 Reconstruction of Convolutional Codes

Since any punctured convolutional code is equivalent to a non punctured convolutional encoder, we will thus focus on the reconstruction of the latter codes. As far as code reconstruction is concerned, it is worth mentioning that the use of punctured codes make it more complex since we have equivalent non punctured codes whose parameters have higher values, for suitable values of $I, k$ and $M$.

It is always possible to reconstruct convolutional codes in offline mode. This is basically not a problem since for most real cases, convolutional encoders do not change very often since they are hardwired (as an example, two convolutional encoders of constraint length of 9 are embedded in the UMTS standard [1]). Consequently we can spend a lot of time to reconstruct them since the work is done just once. However, there are only a very few known cases (most of them are for tactical, military communications like in the Czech army at least during the 90s) where the encoders are randomly generated right before the transmission. The aim is clearly to hinder the code reconstruction strongly, which therefore cannot be performed online. In this latter case, except for very small values of parameters and noise probability, the reconstruction is too much time consuming.

The reconstruction of convolutional codes is a very mathematical stuff and consequently we will not present it here (see [6, 3] for an exhaustive study). For our purposes, it is just necessary to recall the most significant results with respect to convolutional codes reconstruction.

While it is always possible to make the probability of false alarm (*i.e.* to reconstruct a wrong encoder) tends towards zero, the probability of success depends on many factors but the noise parameter has the most significant impact. Beyond 2-3 % the reconstruction will fail unless having a large amount of encoded sequence or/and accepting to spend a lot of time/machine ressources. In most practical cases, the Viterbi decoding itself is likely to fail for a few percent of noise (less than 0.05) long before the reconstruction process does. Expressing the reconstruction probability of success is not easy from a mathematical point of view and we advise the reader to refer to [6, 3]. Experiments have confirmed that the reconstruction is bound to fail as soon as $p > 3\%$ unless spending a lot of time and computing power.

As for the computational complexity of the reconstruction, the general result [6, 3] states that for a $(n, k, N)$-convolutional code, the lower bound is equal to $\mathcal{O}(\alpha \times n^5 \times N^4)$ where $\alpha(p)$ is a quantity which grows exponentially with the noise probability $p$ [3, Section 2.3.2].

To illustrate that general result, Table 1 gives a few experimental results [6, 3] for a few encoders in the case of a noise level of $10^{-2}$ and $2.10^{-2}$ (Additive White Gaussian noise).

| Encoder | Reconstruction time $(p = 10^{-2})$ | Reconstruction time $(p = 2.10^{-2})$ |
|---------|-------------------------------------|---------------------------------------|
| (4, 3, 8) | 7 min 12 sec | Non detected |
| (4, 3, 9) | 6 min 16 sec | Non detected |

Table 1: Example of reconstruction time (on Pentium IV 2.0 Ghz) for two noise levels

As a consequence, considering a rather high level of noise prevents the reconstruction to succeed unless we devote a huge computing time (several hours) at least. We then will choose a noise level ranging from 0.15 to 0.35.

Let us mention that PERSEUS technology considers (and implements) the worst case of communication channel model with respect to the reconstruction problem: the *Additive White Gaussian* model in which the noise is applied uniformly (in other words the noise variable is a random, identically distributed, independent variable). In real communications (for instance satellite communications) the noise occurs by burst and different channel models must be considered (e.g. *Gilbert-Elliot* model [12]).

## 3 Presentation of the PERSEUS Library

The library includes two main files:

- A header file `perseus.h` which contains the parameters settings, new type definitions and function prototypes.

- A function file `perseus.c` which contains the C code of the different functions: random encoder generation, encoding procedure, decoding procedure...

Additionally, different files are also provided with the library:

- A test program `perseus_test.c` which presents how to implement and use the PERSEUS library.

- A makefile to compile the previous test file.

- A documentation file `howto_libperseus.pdf` and a comprehensive description of library structure and functionalities produced from the source code by means of the `doxygen` utility.

The official code repository is located on `code.google.com/p/libperseus`. The current stable version is 1.0.0.

## 3.1 Setting PERSEUS Parameters

PERSEUS parameters are optimally defined in the *perseus.h* file to provide the best trade-off between security and performance. The reader who would desire to modify those parameters must keep in mind that some of them have an impact on the decoding residual error. So any modification should be envisaged only for programmers having a rather good knowledge in convolutional encoding and Viterbi decoding theory [13].

The main parameters are generated randomly during the encoder generation. So only lower $X_{MIN}$ and upper bounds $X_{MAX}$ are set in order to define a value interval $[X_{MIN}; X_{MAX} + X_{MIN}]$.

### 3.1.1 Encoder inputs

The number of encoder inputs is given by (default values $[1; 6]$).

```
#define KMIN_GEN 1
#define KMAX_GEN 5
```

### 3.1.2 Encoder ouputs

The number of encoder outputs is defined by (default values $[5; 11]$).

```
#define NMIN_GEN 5
#define NMAX_GEN 6
```

### 3.1.3 Constraint length (encoder memory)

The size of the encoder memory (which also determines the degree of encoder polynomials) are defined by (default $[20; 30]$).

```
#define MIN_CONT 10
#define MAX_CONT 20
```

### 3.1.4 Puncturing matrix width

The width of the puncturing matrix whose height is defined by the value $N \in [NMIN_{GEN}; NMIN_{GEN} + NMAX_{GEN}]$ (default $[6; 21]$).

```
#define MIN_MATWIDTH 6
#define MAX_MATWIDTH 16
```

The puncturing level is defined by the number of null entries of that matrix. This number is defined as follows

```
1  * Random generation of the puncturing matrix weight */
2    /* (code->mN*code->mMatWidth − nbzero)           */
3    /* where nbzero = (code->mN*code->mMatWidth/8) */
4    nbzero = (int)((float)((code->mN*code->mMatWidth) >> 3));
5    code->mMatDepth = (code->mN*code->mMatWidth) − nbzero;
```

Let us notice that it is possible to adapt the weight of the punctured matrix according to the values of $N$ and `mMatWidth`. For rather large values of their product it is possible to divide by 16 or even 32 to avoid decoding error arising on low memory computers. PERSEUS library 2.x will implement such optimizations along with combinatorial puncturing patterns.

### 3.2 PERSEUS Security Parameters

There is only one parameter which has a direct impact on the PERSEUS security with respect to the encoder reconstruction problem from noisy sequences. This parameter ensures that this problem remains hard in practice requiring a huge supercomputing power during several days or even weeks for a single encoder.

This parameter is defined in the `Gen_Noise_Generator` function located in the *perseus.c* file.

```
1  /* Noise probability generation ([0.15, 0.35]) */
2    aNGen->proba = 15 + (int)(20.0 * alea());
```

A noise probability close to 0.15 will in average require a reconstruction time in days while a probability close to 0.35 will require weeks or even months of computing time.

The reader must be aware that whenever a noise probability close to 0.50 is not possible in the context of PERSEUS. Such a probability relates to cryptography not to noisy communications.

### 3.3 PERSEUS Noise Generator

In PERSEUS library 1.0.0, the random generator is fixed (it will be random from versions 2.x). This generator is a biased stream cipher (combining generator [8] class). It is initialized by a random 102-bit key which fills up the four linear feedback shift registers (LFSR) at time instant $t = 0$. It is worth noticing that the size of the key prevents exhaustive search (to remove the noise by the attacker) only. Hence the only possible approach is to reconstruct the encoder in the context of a noisy communication.

The four LFSR polynomials are defined in the `perseus.h` file as follows:

```
1  /* Noise generator feedback polynomial 1 */
2  #define POLY1 0x47E07L
3  #define MASK1 0x7FFFFL
```

```
4   #define LR1 19
5
6   /* Noise generator feedback polynomial 2 */
7   #define POLY2 0x1772AFL
8   #define MASK2 0x7FFFFFL
9   #define LR2 23
10
11  /* Noise generator feedback polynomial 3 */
12  #define POLY3 0x1C95269L
13  #define MASK3 0x1FFFFFFFL
14  #define LR3 29
15
16  /* Noise generator feedback polynomial 4 */
17  #define POLY4 0x43E98841L
18  #define MASK4 0x7FFFFFFFL
19  #define LR4 31
```

The biased filtering Boolean function which outputs the additive noise to combine with the encoded sequence is then defined by

```
1   /* Noise probability generation ([0.15, 0.35]) */
2   aNGen->proba = 15 + (int)(20.0 * alea());
3
4   /* Boolean filtering function generation */
5   w = 0;
6   aNGen->Bf = (unsigned char *)calloc(16, sizeof(unsigned char));
7   for (w = 0; w < 16; w++)
8     {
9       val = (int)(99.0 * alea());
10      if(val < aNGen->proba) aNGen->Bf[w] = 1;
11    }
```

## 4 Implementation of the PERSEUS Library

Using and implementing the PERSEUS library is almost straightforward and easy (Figure 2). In order to illustrate things, a sample test file `perseus_test.c` is provided with the library [18]. We are going to detail the whole process as it is in the library howto file. Let us mention that since the library uses dynamic Viterbi decoding (which may be memory consumming depending on the instances of PERSEUS parameters, the decoding may fail if you choose to process large amount of data on a computer with limited memory. We strongly advise to split data into chunks of less than 2 Kb. The next version of the library (from versions 2.x) will consider polynomial time decoding anf therefore this limitation will no longer exist.

Let us suppose that the data to protect are stored into the array `data`. John Doe from USA wants to send them to Jean Martin in France in a secure way.

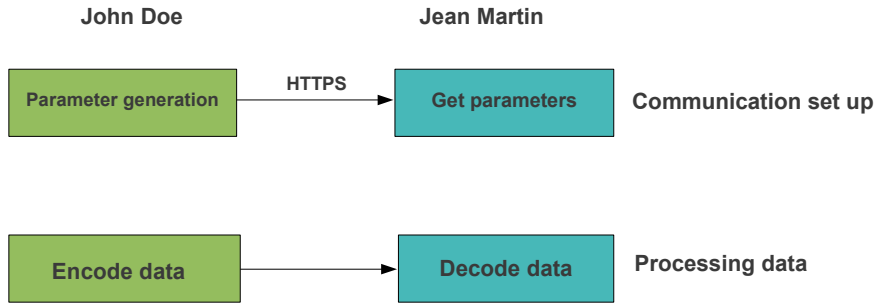

Figure 2: Implementation structure of the PERSEUS library

On John Doe's side, the main steps are (in the following order):

1. First generating the encoder, the noise generator and the noise generator secret key randomly.

```
1     /* Generate the PCC encoder    */
2     Pcc = generateCode();
3     ...
4     /* Noise generator secret key generation */
5     aKey = (INIT_NOISE_GEN *)calloc(1,
6             sizeof(INIT_NOISE_GEN));
7     ...
8     aKey->INIT1 = (unsigned long int)((float)
9                     (0xFFFFFFFFL)*alea());
10    aKey->INIT2 = (unsigned long int)((float)
11                    (0xFFFFFFFFL)*alea());
12    aKey->INIT3 = (unsigned long int)((float)
13                    (0xFFFFFFFFL)*alea());
14    aKey->INIT4 = (unsigned long int)((float)
15                    (0xFFFFFFFFL)*alea());
16
17    /* Noise generator variable allocation    */
18    NGen = (NOISE_GEN *)calloc(1, sizeof(NOISE_GEN));
19    ...
20    /* Noise generator init           */
21    if (!Gen_Noise_Generator(NGen, aKey))
22    {
23     perror("Noise encoder generation on error!");
24     free(NGen);
25     exit(0);
26    }
27    ...
```

2. Sending the secret elements to Jean Martin through a HTTPS session (or

any equivalent secure channel). This part is not played in the *perseus_test.c* file (obvious to implement). The secret elements are the PCC encoder and the noise generator secret key. It consists in three structures (defined in file *perseus.h*)

```c
/* Generic type for Punctured
                Convolutional Code */
typedef struct
  {
  unsigned int mN;
      /* Number of output bits   */
  unsigned int mK;
      /* Number of input bits    */
  unsigned int mM;
      /* Encoder memory size     */
  unsigned long ** mPoly;
      /* Encoder polynomials     */
  unsigned int mMatWidth;
      /* Puncturing matrix width */
  unsigned char * mMatrix;
      /* Encoder puncturing matrix */
  unsigned int mMatDepth;
      /* Puncturing matrix weight */
  } PUNCT_CONC_CODE;

/* Generic type for a noise generator */
typedef struct
  {
  unsigned long int Reg1;
  /* Linear Feedback Shift Register 1 */
  unsigned long int Reg2;
  /* Linear Feedback Shift Register 2 */
  unsigned long int Reg3;
  /* Linear Feedback Shift Register 3 */
  unsigned long int Reg4;
  /* Linear Feedback Shift Register 4 */
  unsigned int L1;
  /* Length of LFSR 1 */
  unsigned int L2;
  /* Length of LFSR 2 */
  unsigned int L3;
  /* Length of LFSR 3 */
  unsigned int L4;
  /* Length of LFSR 4 */
  unsigned char * Bf;
  /* Combining Boolean function */
  unsigned int proba;
  /* Noise probability           */
  } NOISE_GEN;
```

```
45
46         /* Generic type for noise generator
47            secret key   */
48         typedef struct
49          {
50           unsigned long int INIT1;
51           unsigned long int INIT2;
52           unsigned long int INIT3;
53           unsigned long int INIT4;
54          } INIT_NOISE_GEN;
```

3. Encoding the data

```
1          encoded_data_size = 0L;
2          if (!pcc_Code(Pcc, data, data_size, &encoded_data,
3                        &encoded_data_size, NGen, aKey))
4           {
5            perror("Encoding error\n");
6            exit(0);
7           }
8
9          printf("Data after encoding = %s\n", encoded_data);
```

   The PCC encoding includes all basic steps (character to binary encoding,
   the PCC coding itself, data puncturing right after the encoding, the binary
   to hex nibbles encoding, the addition of deterministic noise). The final
   result of the PCC encoding is contained in the array `encoded_data`.

4. John Doe sends the encoded data to Jean Martin.

On Jean Martin's side, the steps are:

1. Reception of the secret elements through a HTTPS session (PCC encoder
   and the noise generator secret key) from John Doe. The three correspond-
   ing data structures (see above John Doe's step 2) are then initialized. This
   part is not played in the *perseus_test.c* file (obvious to implement).

2. Decode data

```
1          dataLength = 0L;
2          if (!pcc_decode(Pcc, NGen, aKey, encoded_data,
3           encoded_data_size, &dataDecoded, &dataLength))
4           {
5            perror("Decoding error\n");
6            exit(1);
7           }
```

The PCC decoding step includes all basic processings (remove the deterministic noise, hex nibble to binary transcoding, data unpuncturing and Viterbi decoding). Encoded data are in the array `dataCoded` while `Decoded data` are contained in the array `dataDecoded`.

## 5 Experimental Results

We have tested our implementation of the Perseus library on a 2 Gb RAM, Intel Core2 Duo CPU P8400 (2.26GHz). Data have been processed by chunks of 1 or 2 Kb. Of course the performance are depending on the random instances of encoders. The main bottleneck remains the dynamic Viterbi decoding which takes most of the processing time (more than 70 % of the total time) and of the available memory. However average performances are rather good. Let us notice that the current release (1.0.0) has not been optimized to preserve the code readibility.

The next version of the Perseus library will consider a polynomial time decoding while requiring a negligible amount of memory.

### 5.1 Perseus Entropy Profile

In order to illustrate the fact that Perseus-protected data may exhibit an entropy profile which is close to that of plain (unprotected) data, we have computed the average entropy per byte on several files (on different Indo-European languages). Table 2 summarizes the results.

| Noise probability | Plain data average entropy | Perseus-protected data | Encrypted data |
|---|---|---|---|
| 5 % | 4.21 | 4.96 | 8.00 |
| 10 % | 4.21 | 6.19 | 8.00 |
| 15 % | 4.21 | 6.46 | 8.00 |
| 20 % | 4.21 | 7.11 | 8.00 |
| 25 % | 4.21 | 7.39 | 8.00 |
| 30 % | 4.21 | 7.45 | 8.00 |
| 35 % | 4.21 | 7.71 | 8.00 |

Table 2: Average entropy profile for plain, Perseus-protected and AES encrypted data

These results clearly show that the entropy profile depends on the noise level (which is quite obvious). Our tests have also confirmed that the more complex the encoder is (in terms of redundancy added) the lower the entropy profile is.

Let us recall that the convolutional code reconstruction is untractable (in reasonable amount of time) as soon as noise probability is higher than a few percents (practically $> 0.02$). So if we want to lower the entropy profile, we can consider noise probability of 5% while preserving the scalable-security provided by the PERSEUS approach.

## 5.2  Secure Programming

Throughout the programming process, the code security was a priority. We have paid a maximal attention to this point. Once the PERSEUS library has been achieved, we have performed code auditing with respect to security.

We have first applied the `Flawfinder` utility [10] which tracks unsecure programming. It helps preventing buffer overflows, heap overflows... by checking the nature and use of common functions. In a second step, we have analyzed how efficiently and correctly the PERSEUS library uses memory. For that purpose, the `Valgrind` utility [19] has been considered.

As a result, the C code of the PERSEUS library complies with the existing rules of secure programming and hence does not introduce weakness or flaws that could be exploited for attack purposes.

# 6  Applications and Implementations

At the present time, a few implementations and application of the PERSEUS technology are known. We hope that new contributors will volunteer to give birth to new ones.

The DFT Technologie company (http://www.dft-techno.com) has decided to provide the industry support to the PERSEUS technology and to help and promote the research and development effort around it.

## 6.1  Firefox Plug-in

This project is managed by Eddy Deligne [17]. He has applied the PERSEUS technology to protect HTTP protocol (GET and POST methods) while using Firefox [5]. This solution is materialized in the form of a C++ Firefox plug-in developed under the triple GPL/LGPL/MPL licences and complying with the specifications of *Mozilla* development, thus allowing the code to be merged to the Firefox engine code directly. This plug-in is available with the corresponding server (Linux, Windows) thus providing an all-round solution (client/server architecture).

At the present time, all Firefox versions 3.x are covered (Windows, Linux, Apple). The new Firefox 4.x should be also protected very soon (many structural changes have occured with this new version thus requiring significant changes in the PERSEUS plug-in).

## 6.2 Andromeda Library: Protecting the Torrent Protocol

Fabien Jobin [2] has developed the ANDROMEDE library[4] which implements the bittorrent protocol in its original version (e.g. without any additional third-party functionality except one devoted to the extension management). In the ANDROMEDE library, the bittorrent traffic is protected by the PERSEUS technology.

## 7 Conclusion and Future Works

The PERSEUS technology intends to propose a new trend in information and communication security. The concept of scalable security should help to make converge the needs for National Security and citizens' natural rights for privacy. This technology preserves the ability of state intelligence agencies to have access to the PERSEUS-protected data. Indeed the noisy encoding layer can always be processed at the price of an offline, time-consuming computing step. Only national security agencies and specialized police departments have such a suitable computing power. But since it requires a lot of time to break this technology, the number of attempts will be limited to process the communication of really bad guys only and not those of any ordinary citizen.

Current research and development activities around PERSEUS technology consider the protection of voice and phone communications as well as file protection:

- development and implementation of VoIP platforms;
- development of Android modules and apps to provide communication protection for various kind of data: voice, sms, mms...
- development of Linux/Windows application to protect files on hard disk.

The main difficulty here lies in the Viterbi decoding which is the most time-consuming part. However our recent research results to develop a new decoding algorithm which has polynomial complexity are more than very promising. This is of nature to speed up the decoding step significantly, thus opening a lot of opportunities with respect to the PERSEUS technology.

Finally, our current work focus on additional plug-ins which enable first to lower the entropy profile of PERSEUS-protected data in order to make it far closer to plain data and second to make their entropy profile and statistical features look like to those of arbitrary data (image files, PDF files...).

## Acknowledgement

I would like to thank Olivier Ferrand for his guru skills with `Valgring` and `Flawfinder` as well as Eddy Deligne for his help to review the library code.

---

[4]In the Greek mythology, Andromede is Perseus' wife.